\documentclass[aps,showpacs,twocolumn]{revtex4-1}
\usepackage{epsfig}
\usepackage{graphicx}
\usepackage{amsmath,amssymb}
\usepackage{float}

\begin{document}

\title{Full-heavy tetraquarks in constituent quark models}

\newcommand*{\NJNU}{Department of Physics, Nanjing Normal University, Nanjing, Jiangsu 210097, China}\affiliation{\NJNU}

\author{Xin Jin}\email[E-mail: ]{181002005@stu.njnu.edu.cn}
\author{Yaoyao Xue}\email[E-mail: ]{1943432785@qq.com}
\author{Hongxia Huang}
\email[E-mail: ]{hxhuang@njnu.edu.cn (Corresponding author)}
\author{Jialun Ping}
\email[E-mail: ]{jlping@njnu.edu.cn (Corresponding author)}
\affiliation{Department of Physics, Nanjing Normal University, Nanjing, Jiangsu 210097, China}

\begin{abstract}
The full-heavy tetraquarks $bb\bar{b}\bar{b}$ and $cc\bar{c}\bar{c}$ are systematically investigated within the chiral quark model
and the quark delocalization color screening model. Two structures, meson-meson and diquark-antidiquark, are considered. For the full-beauty $bb\bar{b}\bar{b}$ systems, there is no any bound state or resonance state in two structures in the chiral quark model, while the wide resonances with masses around 19.25 GeV and the quantum numbers $J^{P}=0^{+}$, $1^{+}$, and $2^{+}$ are possible in the
quark delocalization color screening model. For the full-charm $cc\bar{c}\bar{c}$ systems, the results
are qualitative consistent in two quark models. No bound state can be found in the
meson-meson configuration, while in the diquark-antidiquark configuration there may exist the resonance states, with masses range
between $6.3$ GeV to $7.4$ GeV, and the quantum numbers $J^{P}=0^{+}$, $1^{+}$, and $2^{+}$. And the separation between the diquark
and the antidiquark indicates that these states may be the compact resonance states. All these full-charm resonance states
are consistent with the latest results at LHCb collaboration and are worth searching in the experiments further.
\end{abstract}


\maketitle

\section{Introduction} \label{introduction}

Multiquark states have been proposed in the early stage of quark models, which are invented by Gell-Man and Zweig~\cite{1964_1,1964_2}.
No one paid attention to the multiquark states at that time. In 1977, Jaffe did a calculation of the spectra and dominant decay couplings
of $QQ\bar{Q}\bar{Q}$ mesons, which were named as tetrquark states, in the quark-bag model~\cite{Jaffe:1976ig,Jaffe:1976ih}.
However, the tetraquarks state became a hot topic only after the observation of exotic state $X(3872)$ by Belle collaboration in 2003~\cite{Choi:2003ue}.
Since then, more and more exotic states were observed and proposed, such as $Y(4260)$, $Zc(3900)$, $X(5568)$, and so on. The study of
exotic states can help us to understand the hadron structures and the hadron-hadron interactions better.

The full-heavy tetraquark states $(QQ\bar{Q}\bar{Q}, Q=c,b)$ attracted extensive attention for the last few years, since such states with very
large energy can be accessed experimentally and easily distinguished from other states. In the experiments, the CMS collaboration measured
pair production of $\Upsilon(1S)$ \cite{Khachatryan:2016ydm}. There was also a claim of the existence of a full-bottom tetraquark states
$bb\bar{b}\bar{b}$~\cite{APS2018}, with a global significance of $3.6 \sigma$ and a mass around $18.4$ GeV, almost $500$ MeV below the threshold
of $\Upsilon \Upsilon$. However, the LHCb collaboration
searched for the exotic state $bb\bar{b}\bar{b}$ in the $\Upsilon(1S)\mu^{+}\mu^{-}$ final state~\cite{Aaij:2018zrb}, and no significant results
was found. For full-charm system, $J/\psi$ pair production \cite{Aaij:2011yc,Aaij:2016bqq,Khachatryan:2014iia} and double $c\bar{c}$
production \cite{Abe:2002rb} have also been measured experimentally, which may be helpful in seeking the exotic state $cc\bar{c}\bar{c}$.
Very recently, the LHCb collaboration reported their preliminary results on the observations of full-charm states~\cite{LHCb2020},
a broad structure in the range 6.2 to 6.8 GeV, a narrower structure at 6.9 GeV with significance about 5$\sigma$ are observed,
and there is also a hint for a structure around 7.2 GeV.

In fact, the possible existence of these full-heavy tetraquark states has already been considered in a number of theoretical work.
Some researches pointed out that these full-heavy tetraquark states are stable under strong interaction decays~\cite{Iwasaki:1975pv,KTChao,Heller:1985cb,Lloyd:2003yc,Berezhnoy:2011xn,Debastiani:2017msn,Bai:2016int,Chen:2016jxd,Esposito:2018cwh}.
Iwasaki first proposed that $cc\bar{c}\bar{c}$ is a stable bound state with the mass of about $6.0$ GeV or $6.2$ GeV in 1975~\cite{Iwasaki:1975pv}.
Chao predicted that there existed full-charm diquark-antidiquark states with masses in the range 6.4$\sim$ 6.8 GeV~\cite{KTChao}.
Heller {\em et al.} used the potential energy
coming from the MIT bag model and found that the dimesons $cc\bar{c}\bar{c}$ and $bb\bar{b}\bar{b}$ are bound, and the binding energy are
in the range of $0.16 \sim 0.22$ GeV, which depends on the parameters~\cite{Heller:1985cb}.
Lloyd {\em et al.} used a parametrized Hamiltonian to compute the spectrum of all-charm tetraquark states and obtained several close-lying
bound states~\cite{Lloyd:2003yc}.
Berezhnoy {\em et al.} took diquark and antidiquark as point particles and employed hyperfine interaction between them, the masses
of $cc\bar{c}\bar{c}$ and $bb\bar{b}\bar{b}$ states are under the thresholds for $J=1,2$~\cite{Berezhnoy:2011xn}.
Debastiani {\em et al.} used a non-relativistic model to study the spectroscopy of a tetraquark composed of
$cc\bar{c}\bar{c}$ in a diquark-antidiquark configuration, and they found that the lowest S-wave tetraquarks might be below their thresholds of
spontaneous dissociation into low-lying charmonium pairs~\cite{Debastiani:2017msn}.
Yang Bai {\em et al.} also calculated the mass of $bb\bar{b}\bar{b}$ state in a diffusion Monte Carlo method and found it was around 100 MeV
below the threshold of $\eta_{b}\eta_{b}$~\cite{Bai:2016int}.
Esposito {\em et al.} showed that the energy of
$bb\bar{b}\bar{b}$  tetraquark is $18.8$ GeV, approximately 100 MeV below the threshold~\cite{Esposito:2018cwh}.
Wei Chen {\em et al.} studied the existence of exotic doubly hidden-charm/bottom
tetraquark states made of four heavy quarks in a moment QCD sum rule method and discovered that the masses of the $bb\bar{b}\bar{b}$ tetraquarks
are all below or very close to the thresholds of $\Upsilon(1S)\Upsilon(1S)$ and $\eta_{b}(1S)\eta_{b}(1S)$, except one current of
$J^{PC}=0^{++}$ and the masses of the $cc\bar{c}\bar{c}$ tetraquarks are all above the threshold~\cite{Chen:2016jxd}.
However, many researches disapproved the existence of the full-heavy tetraquark
states~\cite{Wu:2016vtq,Karliner:2016zzc,Hughes:2017xie,Richard:2017vry,Liu:2019zuc,Wang:2019rdo,Chen:2019dvd,Chen:2020lgj,Deng:2020iqw}.
Karliner {\em et al.} estimated the masses of the lowest-lying $cc\bar{c}\bar{c}$ and $bb\bar{b}\bar{b}$ tetraquarks by using the color-magnetic
interaction model, and found they were 6192$\pm 25$ MeV and $18826\pm 25$ MeV, respectively, which were higher than the corresponding
thresholds~\cite{Karliner:2016zzc}.
Hughes {\em et al.} searched for the beauty-fully bound tetraquarks by using the lattice nonrelativistic QCD method, and found no evidence
of a QCD bound tetraquark below the lowest noninteracting thresholds in the channels studied~\cite{Hughes:2017xie}. Mingsheng Liu {\em et al.}
studied the mass spectra of the all-heavy tetraquark systems within a potential model, and suggested that no bound state can be
formed~\cite{Liu:2019zuc}. Chen studied the ground states of the beauty-full and charm-full systems in a nonrelativistic chiral quark model
with the help of Gaussian expansion method~\cite{Chen:2019dvd,Chen:2020lgj}, and
Deng employed three quark models by using the same method~\cite{Deng:2020iqw}, neither of them found the $cc\bar{c}\bar{c}$ and
$bb\bar{b}\bar{b}$ bound states.

Quantum chromodynamics(QCD) is widely accepted as a fundamental theory of strong interaction. However, it is
difficult to study hadron-hadron interactions and multiquark states directly because of nonperturbtive properties of QCD in the low energy region.
Many quark models based on QCD theory have been developed to get physical insights into the multiquark systems. A common approach is the chiral
quark model (ChQM)~\cite{Salamanca1}, in which the constituent quarks interact with each other through colorless Goldstone bosons exchange
in addition to the colorful
one-gluon-exchange and confinement, and the chiral partner $\sigma$ meson-exchange is introduced to give the immediate-range attraction of
nucleon-nucleon interaction. An alternative approach to study the hadron-hadron interactions is the quark delocalization color screening
model (QDCSM), which was developed in the 1990s with the aim of explaining the similarities between nuclear and molecular forces~\cite{Wang:1992wi}.
The quark delocalization and color screening in QDCSM work together to provide the short-range repulsion and the intermediate-range attraction.
Both of these two models can give a good description of the properties of deuteron, nucleon-nucleon and hyperon-nucleon
interactions~\cite{ChenLZ,ChenM}. Recently, QDCSM has been used to study the pentaquarks with hidden strange~\cite{Huang:2018ehi}, hidden-charm
and hidden-bottom~\cite{Huang:2015uda}. Besides, this model was also applied to the tetraquarks composed of $us\bar{d}\bar{b}$ and $ud\bar{s}\bar{b}$
to investigate the existence of $X(5568)$~\cite{Huang_X}. Therefore, it is interesting to extend this model to the full-heavy tetraquarks.
In addition, to check the model dependence of the full-heavy tetraquarks and explore the hadron-hadron interactions in different quark models,
the ChQM is also used in this work.

The structure of this paper is as follows. section II gives a brief introduction of two quark models, and the construction of wave functions.
The numerical results and discussions are given in Section III. The summary is presented in the last section.

\section{Models and wavefunctions}
In this work, we investigate the full-heavy tetraquarks within two quark models: ChQM and QDCSM. Two structures, meson-meson
and diquark-antidiquark, are considered. In this sector, we will introduce these two models and the wave functions of the tetraquarks
for two structures.

\subsection{The chiral quark model (ChQM)}
The ChQM has been successfully applied to describe the properties of hadrons and hadron-hadron interactions~\cite{Salamanca1,Salamanca2},
and has been extended to the study of multiquark states. The model details can be found in Ref.~\cite{Salamanca1,Salamanca2}. We only show
the Hamiltonian of the model here.
\begin{align}
H={}&\sum_{i=1}^{4}( m_i+\frac{p_i^2}{2m_i})-T_{cm}+\sum_{i=1<j}^{4}( V^{CON}_{ij}+V^{OGE}_{ij})
\end{align}
where $T_{cm}$ is the kinetic energy of the center of mass; $V^{CON}_{ij}$ and $V^{OGE}_{ij}$ are the interactions of the confinement and the one-gluon-exchange, respectively. For the full-heavy system, there is no $\sigma$-exchange and the Goldstone boson exchange. The forms of $V^{CON}_{ij}$ and $V^{OGE}_{ij}$ are shown below:
\begin{align}
V^{CON}_{ij}&= -\boldsymbol{\lambda}^c_i \cdot \boldsymbol{\lambda}^c_j (a_c r^2_{ij}+V_{0}) \\
V^{OGE}_{ij}&=\frac{\alpha_s}{4} \boldsymbol{\lambda}^c_i \cdot \boldsymbol{\lambda}^c_j
  [\frac{1}{r_{ij}}-\frac{\pi}{2}\delta(\mathbf{r}_{ij})(\frac{1}{m_{i}^{2}}+\frac{1}{m_{j}^{2}}
  +\frac{4\boldsymbol{\sigma}_{i}\cdot\boldsymbol{\sigma}_{j}}{3m_{i}m_{j}})\notag\\
              {}&-\frac{3}{4m_{i}m_{j}r_{ij}^{3}}S_{ij}]\\
S_{ij}&=\{3\frac{(\boldsymbol{\sigma}_{i}\cdot \boldsymbol{r}_{ij})(\boldsymbol{\sigma}_{j}\cdot \boldsymbol{r}_{ij})}{r_{ij}^{2}}
  -\boldsymbol{\sigma}_{i}\cdot \boldsymbol{\sigma}_{j}\}
\end{align}
where $S_{ij}$ is quark tensor operator; $\alpha_{s}$ is the quark-gluon coupling constant.

\subsection{The quark delocalization color screening model (QDCSM)}
Generally, the Hamiltonian of QDCSM is almost the same as that of ChQM, but with two modifications~\cite{Wang:1992wi}. The one is that there
is no $\sigma$-meson exchange in QDCSM, and another one is that the screened color confinement is used between quark pairs reside in different
clusters, aimed to take into account the QCD effect which has not yet been included in the two-body
confinement and effective one gluon exchange. Since there is no $\sigma$-meson exchange interaction between the heavy quarks ($c$ or $b$), the only difference here is the confinement interaction. The confining potential in QDCSM was modified as follows:
 $$V^{CON}_{ij}=\left\{
 \begin{array}{rcl}
 -\boldsymbol{\lambda}^c_i \cdot \boldsymbol{\lambda}^c_j (a_c r^2_{ij}+V_{0})&&\text{i,j in the same cluster}\\
 - \boldsymbol{\lambda}^c_i \cdot \boldsymbol{\lambda}^c_j a_c\frac{1 - e^{-\mu_{ij} r^2_{ij}}}{\mu_{ij}}&&\text{, otherwise}\\
 \end{array}\right.$$
where $\mu_{ij}$ is the color screening parameter, which is determined by fitting the deuteron properties, $NN$ scattering phase shifts, and
$N\Lambda$ and $N\Sigma$ scattering phase shifts, respectively, with $\mu_{uu}=0.45~$fm$^{-2}$, $\mu_{us}=0.19~$fm$^{-2}$ and $\mu_{ss}=0.08~$fm$^{-2}$, satisfying the relation, $\mu_{us}^{2}=\mu_{uu}\mu_{ss}$~\cite{ChenM}. When extending to the heavy quark case, there is no experimental
data available, so we take it as a adjustable parameter $\mu_{cc}=0.01 \sim 0.001~$fm$^{-2}$ and $\mu_{bb}=0.001 \sim 0.0001~$fm$^{-2}$. We find the results
are insensitive to the value of $\mu_{cc}$ and $\mu_{bb}$. So in the present work, we take $\mu_{cc}=0.01~$fm$^{-2}$ and $\mu_{bb}=0.001~$fm$^{-2}$.

The single particle orbital wave functions in the ordinary quark cluster model are the left and right centered single Gaussian functions:
\begin{eqnarray}
\phi_\alpha(\mathbf{S}_{i})=\left(\frac{1}{\pi
b^2}\right)^{\frac{3}{4}}e^ {-\frac{(\mathbf{r}-\mathbf{S}_i/2)^2}{2b^2}},
 \nonumber\\
\phi_\beta(-\mathbf{S}_{i})=\left(\frac{1}{\pi
b^2}\right)^{\frac{3}{4}}e^ {-\frac{(\mathbf{r}+\mathbf{S}_i/2)^2}{2b^2}}.
\end{eqnarray}
The quark delocalization in QDCSM is realized by writing the single particle orbital wave function as a linear combination of the left and right Gaussians:
\begin{eqnarray}
{\psi}_{\alpha}(\mathbf{S}_{i},\epsilon) &=&
\left({\phi}_{\alpha}(\mathbf{S}_{i})
+\epsilon{\phi}_{\alpha}(-\mathbf{S}_{i})\right)/N(\epsilon),
\nonumber \\
{\psi}_{\beta}(-\mathbf{S}_{i},\epsilon) &=&
\left({\phi}_{\beta}(-\mathbf{S}_{i})
+\epsilon{\phi}_{\beta}(\mathbf{S}_{i})\right)/N(\epsilon),
\nonumber \\
N(\epsilon)&=&\sqrt{1+\epsilon^2+2\epsilon e^{{-S}_i^2/4b^2}}.
\end{eqnarray}
where $\epsilon(\mathbf{S}_i)$ is the delocalization parameter determined by the dynamics of the quark system rather than
adjusted parameters. In this way, the system can choose its most favorable configuration through its own dynamics in a larger
Hilbert space.

The parameters used in our previous work are determined by fitting the mass spectrum of mesons and baryons including light
quarks ($u,~d,~s$), but the meson composed of heavy quarks like $\eta_{b}$($\eta_{c}$) or $\Upsilon$($J/\psi$) do not fit well.
To give the right mass of the mesons we used in this work, we adjust the parameters by fitting the masses of $\eta_{b}$, $\Upsilon$,
$\eta_{c}$ and $J/\psi$. All model parameters used in this work are shown in Table~\ref{parameters}. The calculated masses of the mesons
are $m_{\eta_{b}}=9399$ MeV, $m_{\Upsilon}=9460$ MeV, $m_{\eta_{c}}=2979$ MeV and $m_{J/\psi}=3097$ MeV.
\begin{table}[!htb]
\begin{center}
\caption{Model parameters.}
\renewcommand\arraystretch{1.8}
\begin{tabular}{cccccc}
\hline
  \hline
  $m_b$ (MeV) & $b_{bb}$ (fm) & $a_{c_{bb}}$ (MeV fm$^{-2}$) & $V_{0_{bb}}$ (MeV) & $\alpha_{s_{bb}}$ \\
  5112 & 0.126  & 101 & -40.5 & 0.583 \\
  \hline
  $m_c$ (MeV) & $b_{cc}$ (fm) & $a_{c_{cc}}$ (MeV fm$^{-2}$) & $V_{0_{cc}}$ (MeV) & $\alpha_{s_{cc}}$ \\
  1728 & 0.2  & 101 & -70.5 & 0.518 \\
  \hline   \hline
\end{tabular}
\label{parameters}
 \end{center}
 \end{table}

\subsection{The wave function}
In this work, we use the resonating group method (RGM)~\cite{RGM}, a well-established method for studying
a bound-state or a scattering problem, to calculate the energy of the full-heavy systems. The wave
function of the four-quark system is of the form

\begin{equation}
\Psi=\mathcal{A}[[\psi^{L}\chi^{\sigma}]_{JM}\chi^{f}\chi^{c}].
\end{equation}
where $\psi^{L},\chi^{\sigma},\chi^{f}$ and $\chi^{c}$ are the orbital, spin, flavor and color wave functions, respectively,
which are given below. The symbol $\mathcal{A}$ is the anti-symmetrization operator. For the meson-meson configuration,
$\mathcal{A}$ is defined as
\begin{equation}
\mathcal{A}=1-P_{13}-P_{24}+P_{13}P_{24}.
\end{equation}
For the diquark-antidiquark configuration,
\begin{equation}
\mathcal{A}=1-P_{12}-P_{34}+P_{12}P_{34}.
\end{equation}

\subsubsection{The orbital wave function}
The total orbital wave function is composed of two internal cluster orbital wave functions ($\psi_{1}(\mathbf{R}_{1})$ and
$\psi_{2}(\mathbf{R}_{2})$), and one relative motion wave function ($\chi_{L}(\mathbf{R})$) between two clusters.
\begin{equation}
\psi^{L}=\psi_{1}(\mathbf{R}_{1})\psi_{2}(\mathbf{R}_{2})\chi_{L}(\mathbf{R})
\end{equation}
where $\mathbf{R}_{1}$ and $\mathbf{R}_{2}$ are the internal coordinates for the cluster 1 and cluster 2, respectively. $\mathbf{R}=\mathbf{R}_{1}-\mathbf{R}_{2}$ is the relative coordinate between the two clusters 1 and 2. $\chi_{L}(\mathbf{R})$ is expanded by gaussian bases
\begin{align}
\chi_{L}(\mathbf{R})={}&\frac{1}{\sqrt{4\pi}}(\frac{3}{2\pi b^{2}})\sum_{i=1}^{n}C_{i}\notag\\
                     {}&\times\int \exp [-\frac{3}{4b^{2}}(\mathbf{R}-\mathbf{s}_{i})^{2}]Y_{LM}(\hat{s}_{i})d\hat{s_{i}}
\end{align}
where $\mathbf{s}_{i}$ is the generate coordinate, $n$ is the number of the gaussian bases, which is determined by the stability of the results.
By doing this is expansion, we can simplify the integro-differential equation to an algebraic equation, solve this generalized eigen-equation
to get the energy of the system more easily. The details of solving the RGM equation can be found in Ref~\cite{RGM}.

\subsubsection{The flavor wave function}
The flavor wave function for the full-heavy system is very simple. For the meson-meson structure,
\begin{gather}
\chi_{m00}^{f1}=b\bar{b}b\bar{b}\\
\chi_{m00}^{f2}=c\bar{c}c\bar{c}
\end{gather}
where the superscript of the $\chi$ is the index of the flavor wave function for meson-meson structure, and the subscript
stands for the isospin $I$ and the third component $I_{z}$.

For the diquark-antidiquark structure,
\begin{gather}
\chi_{d00}^{f1}=bb\bar{b}\bar{b}\\
\chi_{d00}^{f2}=cc\bar{c}\bar{c}
\end{gather}
The upper and lower indices are similar to those of the meson-meson structure.

\subsubsection{The spin wave function}
For the spin part, the wave functions for two-body clusters ($Q\bar{Q}$, $QQ$ or $\bar{Q}\bar{Q}$) are
\begin{gather}
  \chi^{1}_{\sigma11}=\alpha\alpha
  \qquad
  \chi^{2}_{\sigma10}=\sqrt{\frac{1}{2}}(\alpha\beta+\beta\alpha) \notag\\
  \chi^{3}_{\sigma1 -1}=\beta\beta
  \qquad
  \chi^{4}_{\sigma00}=\sqrt{\frac{1}{2}}(\alpha\beta-\beta\alpha)
\end{gather}
Then, the total spin wave functions for the four-quark system can be obtained by coupling the wave functions of two clusters.
\begin{align}
\chi^{\sigma 1}_{00}&=\chi^{4}_{\sigma00}\chi^{4}_{\sigma00} \notag\\
\chi^{\sigma 2}_{00}&=\sqrt{\frac{1}{3}}(\chi^{1}_{\sigma11}\chi^{3}_{\sigma1 -1}-\chi^{2}_{\sigma10}\chi^{2}_{\sigma1 0}
+\chi^{3}_{\sigma1-1}\chi^{1}_{\sigma10}) \notag\\
\chi^{\sigma 3}_{11}&=\chi^{4}_{\sigma00}\chi^{1}_{\sigma11} \notag\\
\chi^{\sigma 4}_{11}&=\chi^{1}_{\sigma11}\chi^{4}_{\sigma00} \notag\\
\chi^{\sigma 5}_{11}&=\sqrt{\frac{1}{2}}(\chi^{1}_{\sigma11}\chi^{2}_{\sigma10}-\chi^{2}_{\sigma10}\chi^{1}_{\sigma11})\notag\\
\chi^{\sigma 6}_{22}&=\chi^{1}_{\sigma11}\chi^{1}_{\sigma11}
\end{align}
The spin wave function of two structures is the same.

\subsubsection{The color wave function}
For the meson-meson structure, we give the wave functions for the two-body clusters ($Q\bar{Q}$) first, which are
\begin{gather}
  \chi^{1}_{c[111]}=\sqrt{\frac{1}{3}}(r\bar{r}+g\bar{g}+b\bar{b})\\
  \chi^{2}_{c[21]}=r\bar{b}
  \qquad
  \chi^{3}_{c[21]}=-r\bar{g}\notag\\
  \chi^{4}_{c[21]}=g\bar{b}
  \qquad
  \chi^{5}_{c[21]}=-b\bar{g}\notag\\
  \chi^{6}_{c[21]}=g\bar{r}
  \qquad
  \chi^{7}_{c[21]}=b\bar{r}\notag\\
  \chi^{8}_{c[21]}=\sqrt{\frac{1}{2}}(r\bar{r}-g\bar{g})\notag\\
  \chi^{9}_{c[21]}=\sqrt{\frac{1}{6}}(-r\bar{r}-g\bar{g}+2b\bar{b})
\end{gather}
where the subscript $[111]$ and $[21]$ stand for the color singlet and color octet cluster respectively.

Then, the total color wave functions for the four-quark system with the meson-meson structure can be obtained by
coupling the wave functions of two clusters.
\begin{align}
\chi^{c1}_{m}&=\chi^{1}_{c[111]}\chi^{1}_{c[111]}\\
\chi^{c2}_{m}&=\sqrt{\frac{1}{8}}(\chi^{2}_{c[21]}\chi^{7}_{c[21]}-\chi^{4}_{c[21]}\chi^{5}_{c[21]}-\chi^{3}_{c[21]}\chi^{6}_{c[21]}\notag\\
                              {}&+\chi^{8}_{c[21]}\chi^{8}_{c[21]}-\chi^{6}_{c[21]}\chi^{3}_{c[21]}+\chi^{9}_{c[21]}\chi^{9}_{c[21]}\notag\\
                              {}&-\chi^{5}_{c[21]}\chi^{4}_{c[21]}+\chi^{7}_{c[21]}\chi^{2}_{c[21]})
\end{align}
where $\chi^{c1}_{m}$ and $\chi^{c2}_{m}$ represent the color wave function for the color-singlet channel ($1\times1$) and
the hidden-color channel ($8\times8$), respectively.

For the diquark-antidiquark structure, we firstly give the color wave functions of the diquark clusters,
\begin{gather}
  \chi^{1}_{c[2]}=rr
  \qquad
  \chi^{2}_{c[2]}=\sqrt{\frac{1}{2}}(rg+gr)
  \qquad
  \chi^{3}_{c[2]}=gg\notag\\
  \chi^{4}_{c[2]}=\sqrt{\frac{1}{2}}(rb+br)
  \qquad
  \chi^{5}_{c[2]}=\sqrt{\frac{1}{2}}(gb+bg)\notag\\
  \chi^{6}_{c[2]}=bb
  \qquad
  \chi^{7}_{c[11]}=\sqrt{\frac{1}{2}}(rg-gr)\notag\\
  \chi^{8}_{c[11]}=\sqrt{\frac{1}{2}}(rb-br)
    \qquad
  \chi^{9}_{c[11]}=\sqrt{\frac{1}{2}}(gb-bg)
\end{gather}
and the color wave functions of the antidiquark clusters,
\begin{gather}
  \chi^{1}_{c[22]}=\bar{r}\bar{r}
  \qquad
  \chi^{2}_{c[22]}=\sqrt{\frac{1}{2}}(\bar{r}\bar{g}+\bar{g}\bar{r})
  \qquad
  \chi^{3}_{c[22]}=\bar{g}\bar{g}\notag\\
  \chi^{4}_{c[22]}=\sqrt{\frac{1}{2}}(\bar{r}\bar{b}+\bar{b}\bar{r})
  \qquad
  \chi^{5}_{c[22]}=\sqrt{\frac{1}{2}}(\bar{g}\bar{b}+\bar{b}\bar{g})\notag\\
  \chi^{6}_{c[22]}=\bar{b}\bar{b}
  \qquad
  \chi^{7}_{c[211]}=\sqrt{\frac{1}{2}}(\bar{r}\bar{g}-\bar{g}\bar{r})\notag\\
  \chi^{8}_{c[211]}=\sqrt{\frac{1}{2}}(\bar{r}\bar{b}-\bar{b}\bar{r})
    \qquad
  \chi^{9}_{c[211]}=\sqrt{\frac{1}{2}}(\bar{g}\bar{b}-\bar{b}\bar{g})
\end{gather}

After that, the total wave functions for the four-quark system with the diquark-antidiquark structure are obtained as below,
\begin{align}
\chi^{c1}_{d}&=\sqrt{\frac{1}{6}}[\chi^{1}_{c[2]}\chi^{1}_{c[22]}-\chi^{2}_{c[2]}\chi^{2}_{c[22]}+\chi^{3}_{c[2]}\chi^{3}_{c[22]}\notag\\
               {}&+\chi^{4}_{c[2]}\chi^{4}_{c[22]}-\chi^{5}_{c[2]}\chi^{5}_{c[22]}+\chi^{5}_{c[2]}\chi^{5}_{c[22]}]\\
\chi^{c2}_{d}&=\sqrt{\frac{1}{3}}[\chi^{7}_{c[11]}\chi^{7}_{c[211]}-\chi^{8}_{c[11]}\chi^{8}_{c[211]}+\chi^{9}_{c[11]}\chi^{9}_{c[211]}]
\end{align}

Finally, we can acquire the total wave functions by substituting the wave functions of the orbital, the spin, the flavor
and the color parts into the Eq.(7) according to the given quantum number of the system.

\section{Numerical results and discussions}

In the framework of two quark models, ChQM and QDCSM, we investigate the full-heavy tetraquarks $bb\bar{b}\bar{b}$ and $cc\bar{c}\bar{c}$
in two structures, meson-meson and diquark-antidiquark. The quantum numbers of the teraquarks we study here are $IJ^{P}=00^{+}$, $01^{+}$,
and $02^{+}$.
For the meson-meson structure ($Q\bar{Q}-Q\bar{Q}$), we take into account of two color configurations in the ChQM, which are the color
singlet-singlet ($1\times1$) and color octet-octet ($8\times8$) configurations. Since the QDCSM considers the effect of the hidden color
channel to some extent~\cite{PRC84064001}, only the color singlet-singlet is calculated in this model. For the diquark-antidiquark structure
($QQ-\bar{Q}\bar{Q}$), two color configurations, antitriplet-triplet ($\bar{3}\times3$) and sextet-antisextet ($6\times\bar{6}$),
are considered in both models.
\begin{table}[ht]
\begin{center}
\caption{The energies of $bb\bar{b}\bar{b}$ systems with meson-meson structure in ChQM and QDCSM (unit: MeV).}
\begin{tabular}{ccccccccc}
  \hline
   \hline
  & & & &\multicolumn{2}{c}{ChQM}& \multicolumn{2}{c}{QDCSM} \\
   \hline
  $IJ^{P}$ & $[\chi^{\sigma_{i}}\chi^{f_{j}}\chi^{c_{k}}] $& Channel & $E_{th}$ & $E_{sc}$  & $E_{cc2}$ & $E_{sc}$ &$E_{cc1}$ \\
  \hline
  $00^{+}$ & $\chi^{\sigma 1}_{00} \chi_{m00}^{f1} \chi^{c1}_{m}$ & $\eta_{b}\eta_{b}$ & 18799 & 18803 & 18803  & 18803 & 18803 \\
           & $\chi^{\sigma 2}_{00} \chi_{m00}^{f1} \chi^{c1}_{m}$ & $\Upsilon\Upsilon$ & 18920 & 18925 &        & 18925 &     \\
           & $\chi^{\sigma 1}_{00} \chi_{m00}^{f1} \chi^{c2}_{m}$ & $\eta_{b8}\eta_{b8}$ &     & 19974 &        &       &     \\
           & $\chi^{\sigma 2}_{00} \chi_{m00}^{f1} \chi^{c2}_{m}$ & $\Upsilon_{8}\Upsilon_{8}$ &       & 20041  &       &     \\
  \hline
  $01^{+}$ & $\chi^{\sigma 3}_{11} \chi_{m00}^{f1} \chi^{c1}_{m}$ & $\eta_{b}\Upsilon$ & 18860 & 18864 & 18864  & 18864 & 18864 \\
           & $\chi^{\sigma 4}_{11} \chi_{m00}^{f1} \chi^{c1}_{m}$ & $\Upsilon\eta_{b}$ & 18860 & 18864 &        & 18864 &  \\
           & $\chi^{\sigma 3}_{11} \chi_{m00}^{f1} \chi^{c2}_{m}$ & $\eta_{b8}\Upsilon_{8}$ &  & 19926 &        &       &  \\
           & $\chi^{\sigma 4}_{11} \chi_{m00}^{f1} \chi^{c2}_{m}$ & $\Upsilon_{8}\eta_{b8}$ &  & 19926 &        &       &  \\
  \hline
  $02^{+}$ & $\chi^{\sigma 6}_{11} \chi_{m00}^{f1} \chi^{c1}_{m}$ & $\Upsilon\Upsilon$ &  18920 & 18925 &       & 18925 & 18925   \\
           & $\chi^{\sigma 6}_{11} \chi_{m00}^{f1} \chi^{c2}_{m}$ & $\Upsilon_{8}\Upsilon_{8}$ & & 19923 &      &   & \\
  \hline \hline
\end{tabular}
\label{tab2}

\caption{The energies of $bb\bar{b}\bar{b}$ systems with diquark-antidiquark structure in ChQM and QDCSM (unit: MeV).}
\begin{tabular}{ccccccc}
  \hline
   \hline
   & & &\multicolumn{2}{c}{ChQM}& \multicolumn{2}{c}{QDCSM} \\
   \hline
  $IJ^{P}$ & $[\chi^{\sigma_{i}}\chi^{f_{j}}\chi^{c_{k}}] $ & $E_{th}$ & $E_{sc}$ &$E_{cc2}$ & $E_{sc}$ &$E_{cc2}$ \\
  \hline
  $00^{+}$ & $\chi^{\sigma 1}_{00} \chi_{d00}^{f1} \chi^{c1}_{d}$ & ~18799~ & ~20134~ & ~19466~ & ~19281~ & ~19237~  \\
           & $\chi^{\sigma 2}_{00} \chi_{d00}^{f1} \chi^{c2}_{d}$ &  & 19466 & &19256&   \\
  \hline
  $01^{+}$ & $\chi^{\sigma 5}_{11} \chi_{d00}^{f1} \chi^{c2}_{d}$ & 18860 & 19467 & &19264 & \\
  \hline
  $02^{+}$ & $\chi^{\sigma 6}_{11} \chi_{d00}^{f1} \chi^{c2}_{d}$ & 18920 & 19471 &  &19279 & \\
  \hline \hline
\end{tabular}
\label{tab3}
\end{center}
\end{table}

To find out if there is any bound state in such full-heavy tetraquark systems, we do a dynamic bound-state calculation here.
The single-channel calculation, as well as the channel-coupling are carried out.
All the results for $bb\bar{b}\bar{b}$ and $cc\bar{c}\bar{c}$ systems in two structures are listed in Tables~\ref{tab2}-\ref{tab6}, respectively.
In the tables, the second column shows the combination in spin ($\chi^{\sigma_{i}}$), flavor ($\chi^{f_{j}}$), and
color ($\chi^{c_{k}}$) degrees of freedom for each channel.
The columns headed with $E_{th}$ denotes the theoretical threshold of each channel and $E_{sc}$ represents the lowest energies in the
single channel calculation.
For meson-meson structure, $E_{cc1}$ and $E_{cc2}$ denote the lowest energies of the coupling of the color-singlet channels and the coupling
of all channels, respectively. An additional column headed with "Channel"
denotes the physical contents of the channel. For diquark-antidiquark structure, $E_{cc}$ means the energies of the coupling of all channels.
All the general features of the calculated results are as follows.

\subsection{Full-beauty tetraquarks $bb\bar{b}\bar{b}$}
For the tetraquarks composed of $bb\bar{b}\bar{b}$, the energies of the possible quantum numbers with two structures in both ChQM and
QDCSM are listed in Tables~\ref{tab2} and \ref{tab3}, respectively. The $\eta_{b8}\eta_{b8}$ in the Tables~\ref{tab2} represents
the molecular state $\eta_{b}\eta_{b}$ with the color octet-octet ($8\times8$) configuration. For meson-meson structure, the energies of
every single channel are above the corresponding theoretical threshold in both two models.
The channel-coupling effect is very small and cannot help much, and the energies are still higher than the theoretical thresholds, which means
that there is no any bound states with the meson-meson structure in both two models. Besides, we also find the results in both two models are
almost the same, this is because that the quarks are too heavy to run, resulting in the value of the quark delocalization parameter $\epsilon$
in QDCSM is close to 0. The color screening parameter in QDCSM is also very small because of the heavy quarks, which makes the difference of
the confinement between two models is very small. So both the effect of the quark delocalization and the color screening in QDCSM is very small
in such full-heavy system with meson-meson structure, and the $\sigma$ meson exchange is also inoperative in ChQM, which make the coincident
results of two models.

For the diquark-antidiquark structure, the energies of both ChQM and QDCSM are shown in Tables~\ref{tab3}, from which we can see that the
energies of this configuration are higher than that of meson-meson configuration. Besides, the energies in QDCSM are generally lower than
that in ChQM. Since the color symmetry of the diquark and antidiquark are color octet, the color screening will make the quark delocalization
work in QDCSM, which leads to lower energy in this model. However, the energy of every single channel is still above the corresponding
theoretical threshold. Moreover, the effect of the channel-coupling is also very small in both two models. So none of these states listed in
Tables~\ref{tab3} is a bound state. However, it is possible for them to be resonance states, because the colorful subclusters diquark ($QQ$)
and antidiquark ($\bar{Q}\bar{Q}$) cannot fall apart directly due to the color confinement. To check the possibility, we carry out an adiabatic
calculation of the effective potentials for the $bb\bar{b}\bar{b}$ system with diquark-antidiquark structure, the results of which are shown in Fig. 1.
\begin{figure}[ht]
\begin{center}
\epsfxsize=3.5in \epsfbox{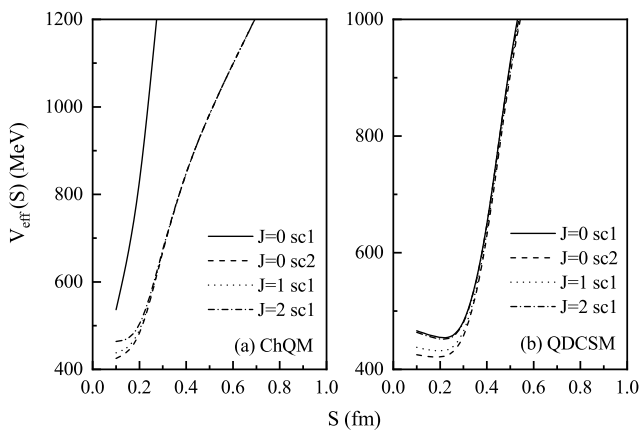} \vspace{-0.3in}

\caption{The effective potentials for the diquark-antidiquark $bb\bar{b}\bar{b}$ systems in two quark models.} \label{pic1}
\end{center}
\end{figure}

From the Fig.~1(a) we can see that the effective potential of each channel is increasing when the two subclusters fall apart, which means
that the diquark and antidiquark tend to clump together without hinderance. This behavior indicates that the odds for the states being
diquark-antidiquark configuration,
meson-meson configuration, or other configurations are the same. Moreover, from the Tables~\ref{tab2} and \ref{tab3}, the energy of the each channel
with the diquark-antidiquark structure is higher than the one of the meson-meson structure. So the state prefers to be two free mesons.
Therefore, none of these state is a observable resonance state in ChQM. It is different in QDCSM (see Fig~1(b)), where the energy of the state
will rise when the two subclusters are too close, so there is a hinderance for the states change structure to meson-meson even if the energy of
the state is lower in meson-meson structure. It is possible to form a wide resonance. The resonance energies are 19237 MeV for $J^P=0^+$, 19264 MeV
for $J^P=1^+$, 19279 MeV for $J^P=2^+$, respectively.

The effective potentials for the meson-meson $bb\bar{b}\bar{b}$ systems are also calculated and shown in Fig.~2 and 3, corresponding to the
color singlet channels and the hidden color channels respectively. From Fig.~2 we find that the effective potentials for the color singlet channels
are all repulsive except for the $IJ^{P}=00^{+}$ $\Upsilon\Upsilon$ channel, which has very weak attractions. That's why we cannot obtain any bound
state in the dynamical calculation.
Fig.~3 only gives the potential for the hidden color channels in ChQM, because the QDCSM considers the effect of the hidden color channel to
some extent, only the color singlet channels are calculated in this model.
It is obvious in Fig.~3 that the behavior of the potential for the hidden color channels is similar to that of the diquark-antidiquark configuration,
where the effective potential of each channel is increasing when the two meson subclusters fall apart. So there is no any observable resonance
state for the meson-meson $bb\bar{b}\bar{b}$ systems.
\begin{figure}[h]
\begin{center}
\epsfxsize=3.5in \epsfbox{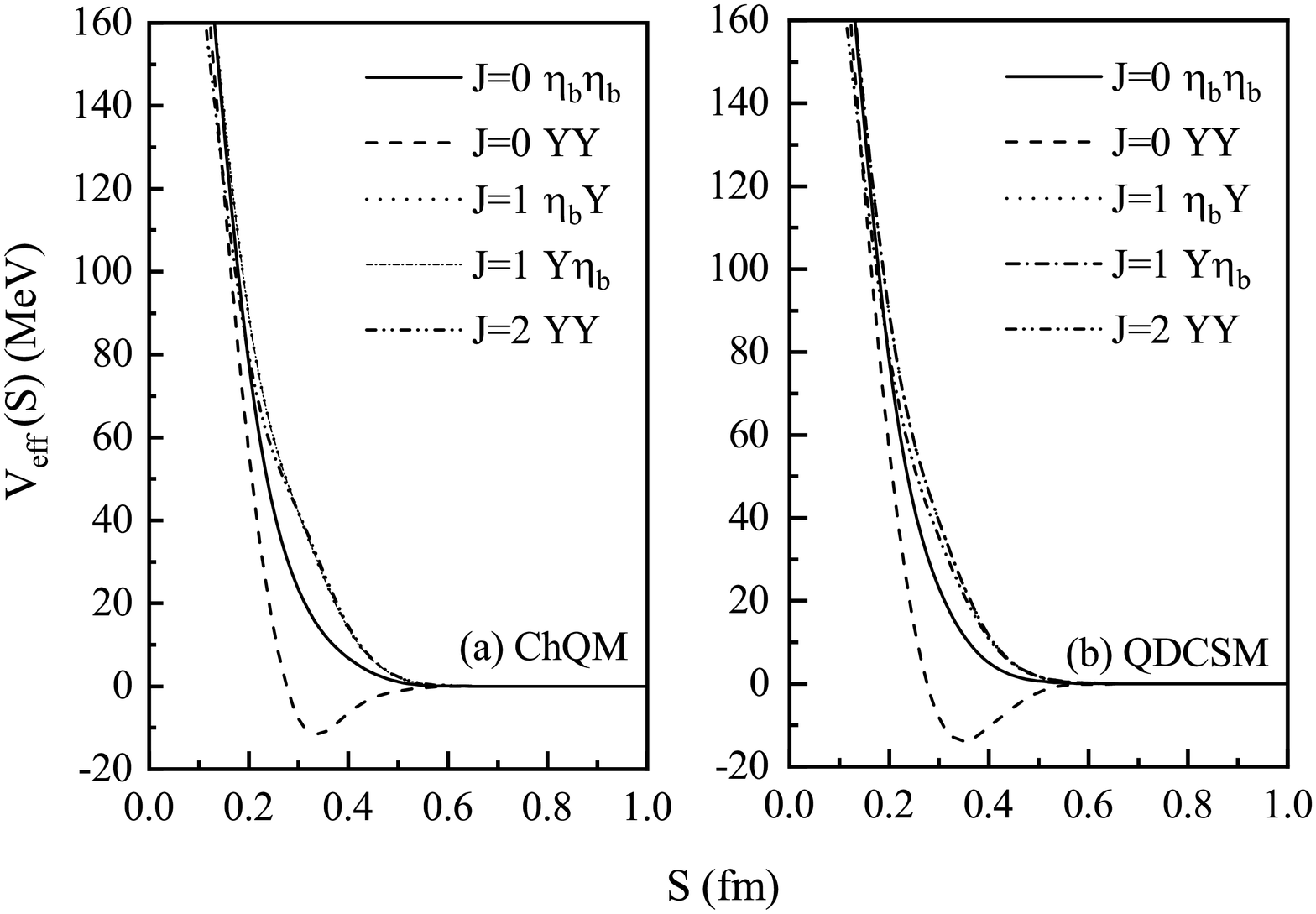} \vspace{-0.3in}

\caption{The effective potentials of the color singlet channels for the meson-meson $bb\bar{b}\bar{b}$ systems in two quark models.} \label{pic2}
\end{center}
\end{figure}

\begin{figure}[h]
\begin{center}
\epsfxsize=3.5in \epsfbox{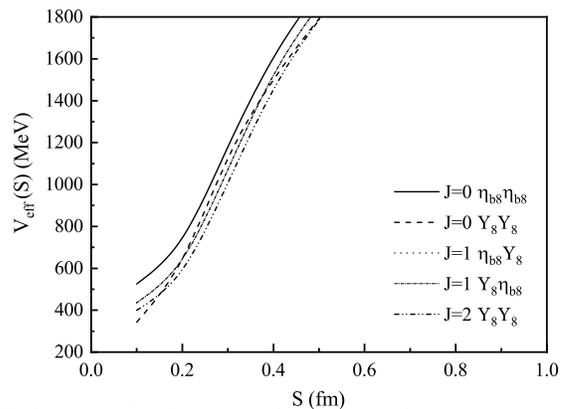} \vspace{-0.45in}

\caption{The effective potentials of the hidden color channels for the meson-meson $bb\bar{b}\bar{b}$ systems in ChQM.} \label{pic3}
\end{center}
\end{figure}

To investigate the interaction between two subclusters, we continue to study the contribution of each interaction term to the energy of the system.
Here, to save space, we take the results of the $IJ^{P}=00^{+}$ $\Upsilon\Upsilon$ channel in meson-meson structure as an example.
The total effective potential, as well as each interaction term to the effective potential, including the kinetic energy ($V_{VK}$),
the confinement ($V_{CON}$), the Coulomb interaction ($V_{Coul}$) and the color-magnetic interaction ($V_{CMI}$), are shown in Fig.~4.
We can see that the kinetic energy term in both ChQM and QDCSM provides attractive interactions, and the attraction in QDCSM is stronger
than the one in ChQM. In the ChQM, the confinement does not contribute to the effective potential between $b\bar{b}$ and $b\bar{b}$,
but it provides a slight repulsion in QDCSM. Both the Coulomb term and the color-magnetic term provide repulsion, which decrease the
attraction of the kinetic energy term, and make the total weak attraction of this channel.

\begin{figure}[ht]
\begin{center}
\epsfxsize=3.5in \epsfbox{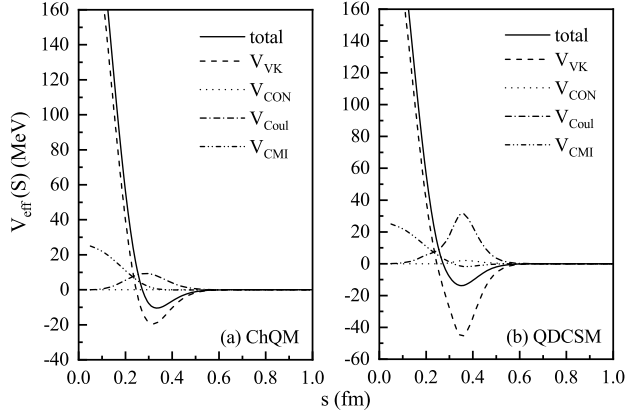} \vspace{-0.3in}

\caption{The contributions to the effective potential of the $IJ^{P}=00^{+}$ $\Upsilon\Upsilon$ channel from various terms of interactions in the ChQM and QDCSM.} \label{pic4}
\end{center}
\end{figure}

\subsection{Full-charm tetraquarks $cc\bar{c}\bar{c}$}
The full-charm systems $cc\bar{c}\bar{c}$ are investigated here and the energies in two quark models with two configurations are listed in
Table~\ref{tab4} and \ref{tab6}, respectively. The results are similar to the full-beauty systems $bb\bar{b}\bar{b}$.
For meson-meson structure, the energy of each single channel is above the corresponding theoretical threshold in both two models.
The energy is almost unchanged by the channel-coupling calculation, indicating that the effect of the channel-coupling is very small.
This is because the mass difference between each channel is very large. So there is no any bound state with the meson-meson structure
in both models.
\begin{table}[ht]
\begin{center}
\caption{The energies of $cc\bar{c}\bar{c}$ systems with meson-meson structure in ChQM and QDCSM (unit: MeV).}
\begin{tabular}{ccccccccc}
  \hline
   \hline
  & & & &\multicolumn{3}{c}{ChQM}& \multicolumn{2}{c}{QDCSM} \\
   \hline
  $IJ^{P}$ & $[\chi^{\sigma_{i}}\chi^{f_{j}}\chi^{c_{k}}] $& Channel & $E_{th}$ & $E_{sc}$  & $E_{cc2}$ & $E_{sc}$ &$E_{cc1}$ \\
  \hline
  $00^{+}$ & $\chi^{\sigma 1}_{00} \chi_{m00}^{f2} \chi^{c1}_{m}$ &$\eta_{c}\eta_{c}$& 5958 & 5961 &  5961 & 5961 & 5961 \\
           & $\chi^{\sigma 2}_{00} \chi_{m00}^{f2} \chi^{c1}_{m}$ & $J/\psi J/\psi$  & 6195 & 6197 &       & 6197 &  \\
           & $\chi^{\sigma 1}_{00} \chi_{m00}^{f2} \chi^{c2}_{m}$ &$\eta_{c8}\eta_{c8}$&    & 6626 &       &      & \\
           & $\chi^{\sigma 2}_{00} \chi_{m00}^{f2} \chi^{c2}_{m}$& $J/\psi_{8} J/\psi_{8}$ & & 6570 &      &      & \\
  \hline
  $01^{+}$ & $\chi^{\sigma 3}_{11} \chi_{m00}^{f2} \chi^{c1}_{m}$ &$\eta_{c}J/\psi$& 6076 & 6079 & 6079 & 6079 & 6079 \\
           & $\chi^{\sigma 4}_{11} \chi_{m00}^{f2} \chi^{c1}_{m}$ &$J/\psi\eta_{c}$& 6076 & 6079 &      & 6079 &  \\
           & $\chi^{\sigma 3}_{11} \chi_{m00}^{f2} \chi^{c2}_{m}$ &$\eta_{c8}J/\psi_{8}$& & 6379 &      &      &  \\
           & $\chi^{\sigma 4}_{11} \chi_{m00}^{f2} \chi^{c2}_{m}$ &$J/\psi_{8}\eta_{c8}$& & 6379 &      &      &  \\
  \hline
  $02^{+}$ & $\chi^{\sigma 6}_{11} \chi_{m00}^{f2} \chi^{c1}_{m}$ &$J/\psi J/\psi$ & 6195 & 6197 & 6197 &  6197 &  \\
           & $\chi^{\sigma 6}_{11} \chi_{m00}^{f2} \chi^{c2}_{m}$ &$J/\psi_{8} J/\psi_{8}$ &  & 6602 &   & &  \\
  \hline \hline
\end{tabular}
\label{tab4}

\caption{The energies of $cc\bar{c}\bar{c}$ systems with diquark-antidiquark structure in ChQM and QDCSM (unit: MeV).\label{tab5}}
\begin{tabular}{ccccccc}
  \hline
   \hline
   & & &\multicolumn{2}{c}{ChQM}& \multicolumn{2}{c}{QDCSM} \\
   \hline
  $IJ^{P}$ & $[\chi^{\sigma_{i}}\chi^{f_{j}}\chi^{c_{k}}] $ & $E_{th}$ & $E_{sc}$ &$E_{cc1}$ & $E_{sc}$ &$E_{cc1}$ \\
  \hline
  $00^{+}$ & $\chi^{\sigma 1}_{00} \chi_{d00}^{f2} \chi^{c1}_{d}$ & ~5958~ & ~6669~ & ~6451~ & ~6405~ & ~6314~  \\
           & $\chi^{\sigma 2}_{00} \chi_{d00}^{f2} \chi^{c2}_{d}$ &  & 6466 & &6358&   \\
  \hline
  $01^{+}$ & $\chi^{\sigma 5}_{11} \chi_{d00}^{f2} \chi^{c2}_{d}$ & 6076 & 6479 & &6375 & \\
  \hline
  $02^{+}$ & $\chi^{\sigma 6}_{11} \chi_{d00}^{f2} \chi^{c2}_{d}$ & 6195 & 6505 &  &6407 & \\
  \hline  \hline
\end{tabular}
\end{center}
\end{table}

To study the interaction between two mesons, we also carry out the adiabatic calculation of the effective potentials for the $cc\bar{c}\bar{c}$
systems, and the results are similar to those of the $bb\bar{b}\bar{b}$ systems. For the color singlet channels, the effective potentials
in both quark models are shown in Fig. 5(a) and (b), respectively. In ChQM, although the effective potential of the $IJ=00$ $J/\psi J/\psi$
is attractive, the attraction is very weak. The potentials of other four channels are all repulsive. So none of these color singlet state is
bound in the dynamical calculation. In QDCSM, although the attraction of the $IJ=00$ $J/\psi J/\psi$ channel is a little stronger than
the one in ChQM, it is still not large enough to form a bound state. Besides, the potential of the $IJ=02$ $J/\psi J/\psi$ channel is attractive too,
but it is very weak. The potential of other three channels are all repulsive. Therefore, there is still no any bound state in QDCSM for the
meson-meson structure. Besides, to investigate if there is any resonance state, we also calculate the effective potential of the hidden color
channels in ChQM, which are shown in Fig.~6. Clearly, the behavior of the potential is similar to that of the full-beauty systems,
where the effective potential of each channel is increasing when the two meson subclusters fall apart. So there is no any observable resonance
state for the $cc\bar{c}\bar{c}$ systems in meson-meson configuration.
\begin{figure}[ht]
\begin{center}
\epsfxsize=3.5in \epsfbox{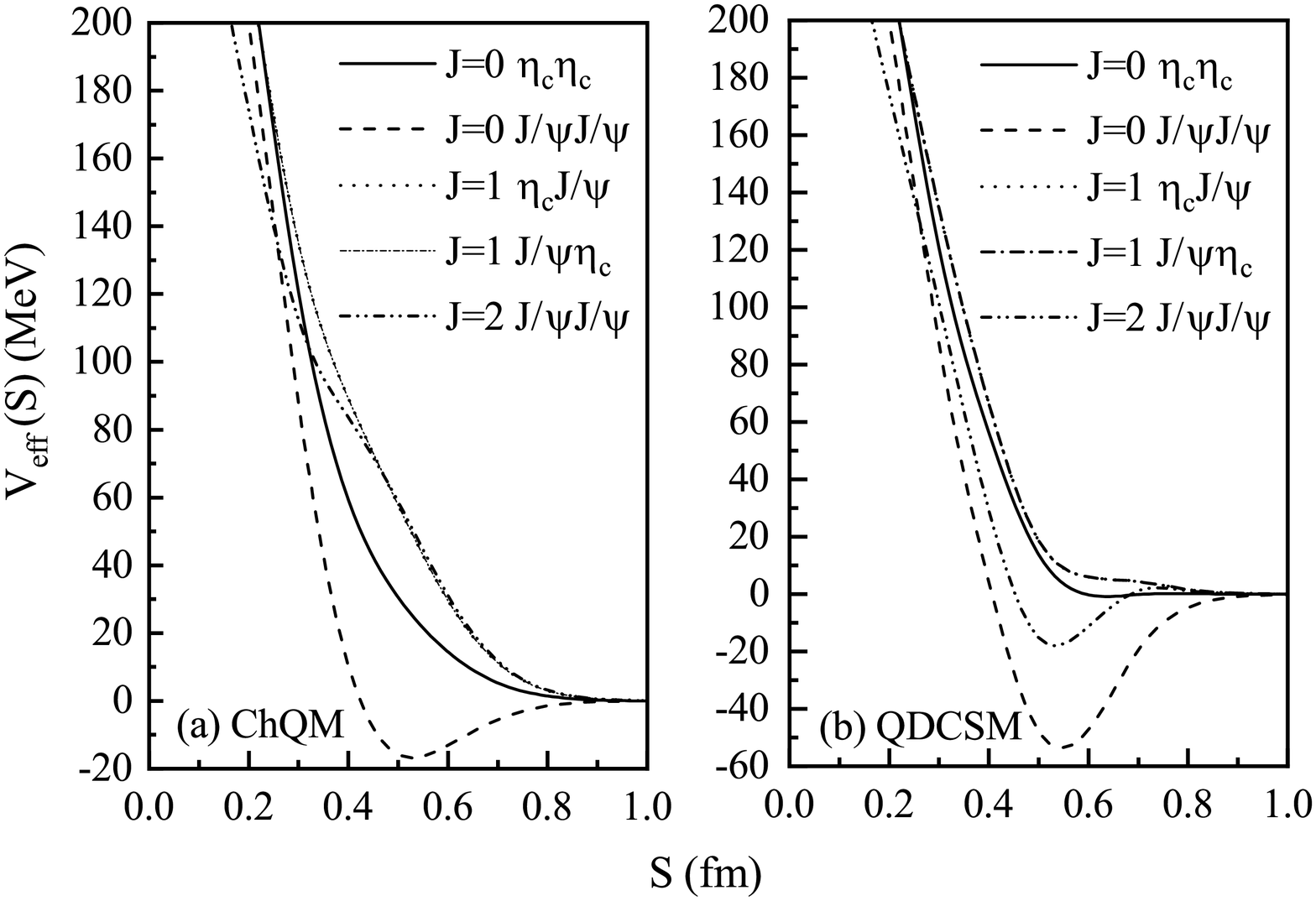} \vspace{-0.3in}

\caption{The effective potentials of the color singlet channels for the meson-meson $cc\bar{c}\bar{c}$ systems in two quark models.} \label{pic5}
\end{center}
\end{figure}

\begin{figure}[ht]
\begin{center}
\epsfxsize=3.5in \epsfbox{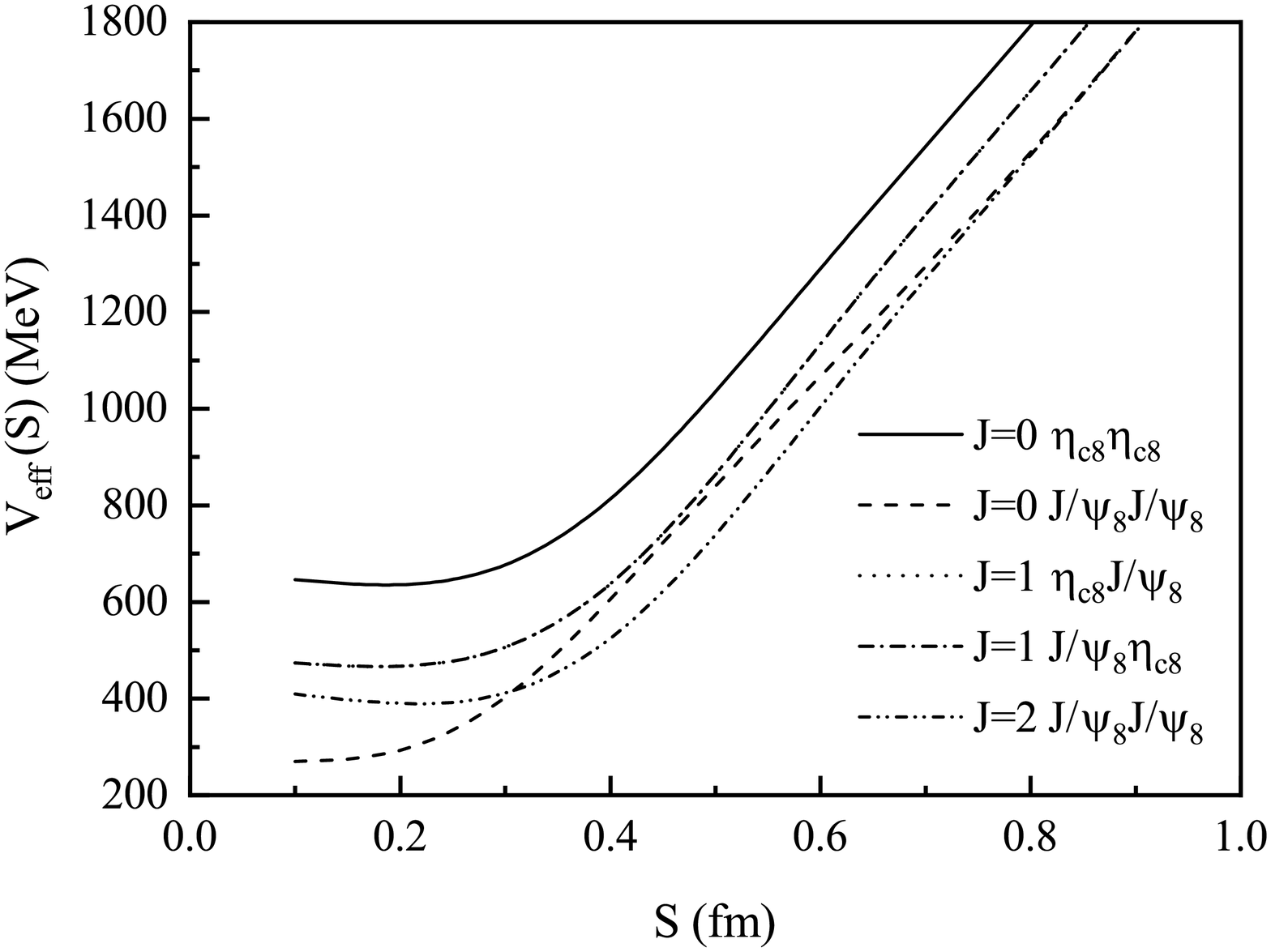} \vspace{-0.3in}

\caption{The effective potentials of the hidden color channels for the meson-meson $cc\bar{c}\bar{c}$ systems in ChQM.} \label{pic6}
\end{center}
\end{figure}

Towards to the diquark-antidiquark configuration, the energies of the systems in both ChQM and QDCSM are listed in Table~\ref{tab5}.
It is clear in Table~\ref{tab5} that the energy of every single channel is above the theoretical threshold of the corresponding channel in
both models. The channel-coupling pushes the lowest energy down a little, but the effect is not large enough to lower the energy below the
threshold. So there is no any bound state in this diquark-antidiquark structure, which is similar to the results of the full beauty systems.
To check that if there is any resonance state, we also carry out the adiabatic calculation of the effective potentials for the
$cc\bar{c}\bar{c}$ systems, as we do for the $bb\bar{b}\bar{b}$ systems.

\begin{figure}[ht]
\begin{center}
\epsfxsize=3.5in \epsfbox{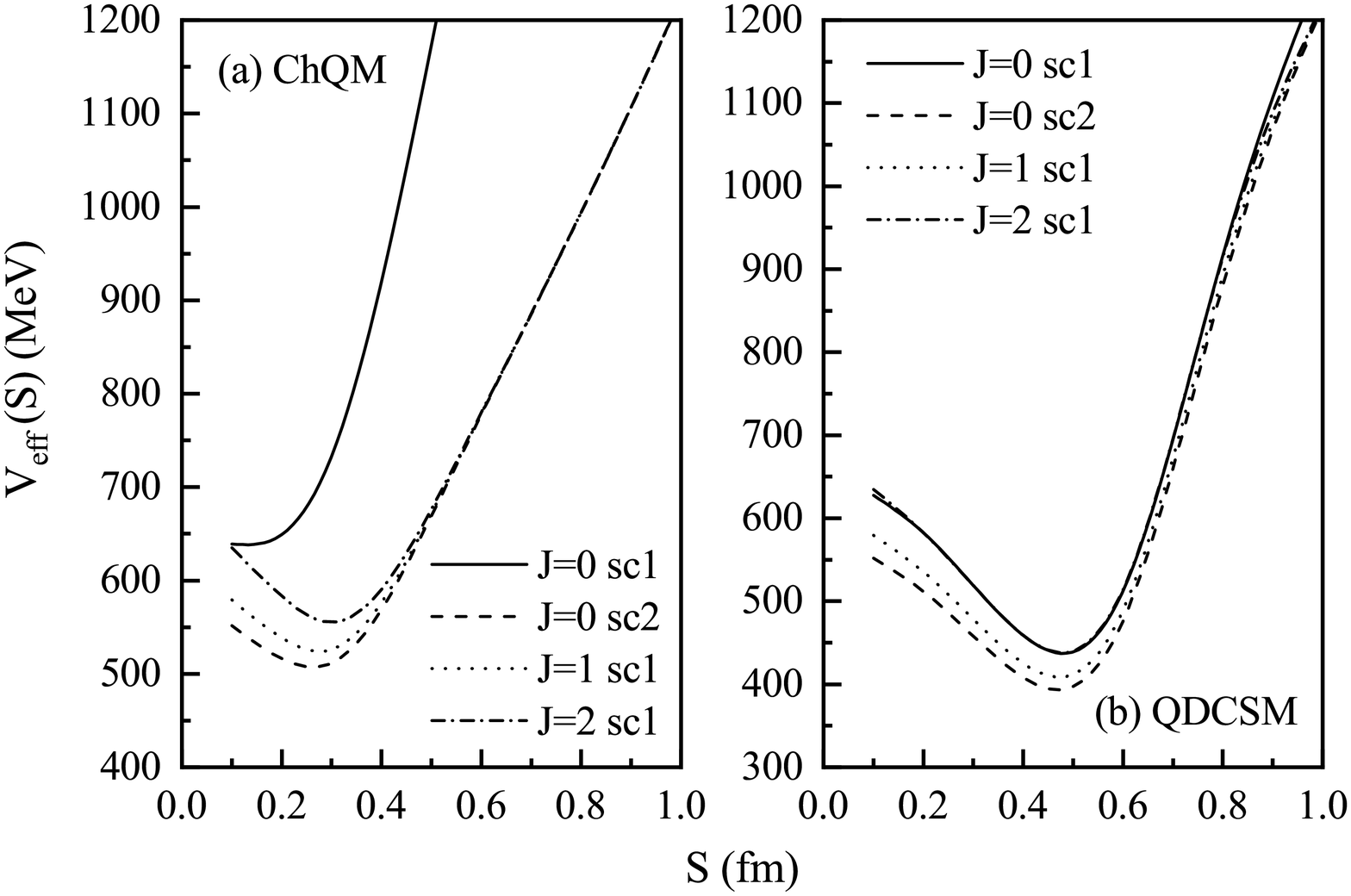} \vspace{-0.3in}

\caption{The effective potentials for the diquark-antidiquark $cc\bar{c}\bar{c}$ systems in two quark models.} \label{pic7}
\end{center}
\end{figure}

The effective potentials for the $cc\bar{c}\bar{c}$ system in diquark-antidiquark structure are shown in Fig.7, from which we find the behavior
of the potential is different from that of the diquark-antidiquark $bb\bar{b}\bar{b}$ systems. In ChQM, the minimum potential of each channel
(except the first channel of $IJ=00$) appears at the separation of $0.3$ fm, which indicates that two subclusters are not willing to huddle
together or fall apart. Besides, the energy of each channel is higher than the corresponding threshold according to Table~\ref{tab5}.
So each state is possible to be a resonance state. We also find that the channel coupling of the $IJ=00$ system makes the energy a little lower.
In QDCSM, the results are similar. The minimum potential of each channel appears at the separation of $0.5$ fm, and the energy of each channel
is higher than the corresponding threshold. So it is also possible for each channel to be a resonance state in QDCSM. From the Table~\ref{tab5},
we can see that the resonance energy in QDCSM is lower than the one in ChQM. The lowest resonance energies in QDCSM are $6314$ MeV for $IJ=00$,
$6375$ MeV for $IJ=01$, $6407$ MeV for $IJ=02$, and those in ChQM are $6451$ MeV for $IJ=00$,  $6479$ MeV for $IJ=01$, $6505$ MeV for $IJ=02$.
Besides, the minimum potential appears at the separation of $0.3$ fm in ChQM and $0.5$ fm in QDCSM, indicating that these two subclusters are
close to each other. Therefore, these resonance states may be the compact resonance states.
\begin{table}[!htb]
\begin{center}
\caption{The energies of $cc\bar{c}\bar{c}$ excited states with the diquark-antidiquark structure in the ChQM and QDCSM.}
\begin{tabular}{cccccccccc}
  \hline
   \hline
   & \multicolumn{4}{c}{ChQM}& \multicolumn{5}{c}{QDCSM} \\
   \hline
  $IJ^{P}$ & \multicolumn{4}{c}{$E$ (MeV)}& \multicolumn{5}{c}{$E$ (MeV)}  \\
  \hline
  $00^{+}$ & 6451 & 6678 & 6923 & 7316 ~&~ 6314 & 6446 & 6815 & 6970 & 7209 \\
  \hline
  $01^{+}$ & 6479 & 6924 & 7318 &      & ~ 6375 & 6885 & 7225 &  \\
  \hline
  $02^{+}$ & 6505 & 6933 & 7323 &      & ~ 6407 & 6924 & 7184 &   \\
  \hline
\end{tabular}
\label{tab6}
\end{center}
 \end{table}

To find the candidates of the exotic states which are reported recently by LHCb collaboration, the energies of the excited states
with diquark-antidiquark structure in two models are also shown in Table~\ref{tab6}. We can see that there are several resonance states with energies
between 6.3$\sim$ 7.4 GeV, and all the states reported by LHCb collaboration can find their candidates as diquark-antidiquark states in our calculation.
To confirm whether they are resonances or not,
the study of the scattering process of the corresponding open channels is needed, which is our future work.

\section{Summary}
In this work, we systematically investigate the low-lying full-heavy systems $bb\bar{b}\bar{b}$ and $cc\bar{c}\bar{c}$ in two quark models:
ChQM and QDCSM. Two structures, meson-meson and diquark-antidiquark, are considered. The dynamic bound-state calculation is carried out to
search for any bound state in the full-heavy systems. To explore the effect of the multi-channel coupling, both the single channel and the
channel coupling calculation are performed. Meanwhile, an adiabatic calculation of the effective potentials is added to study the interactions
of the systems and to find any resonance state.

The numerical results show that: (1) For the full-beauty $bb\bar{b}\bar{b}$ systems, there is no any bound state or resonance state in
two structures in ChQM. While in QDCSM, the wide resonances are possible in the diquark-antidiquark structure, and the resonance energies are 19237 MeV
for $J^P=0^+$, 19264 MeV for $J^P=1^+$, 19279 MeV for $J^P=2^+$.
While for the full-charm $cc\bar{c}\bar{c}$ systems, the resonance states are possible in the diquark-antidiquark
configuration, the lowest energies of which are $6314\sim 6451$ MeV for $IJ=00$, $6375\sim 6479$ MeV for $IJ=01$, and
$6407\sim 6505$ MeV for $IJ=02$. The separation between the diquark and the antidiquark indicates that these states may be compact
resonance states. Besides, we also find several excited resonance states with diquark-antidiquark structure in two models with energies
between 6.3$\sim$ 7.4 GeV. All the states reported recently by LHCb collaboration can find their candidates as diquark-antidiquark states
in our calculation.
(2) The effective potentials show that for most channels, the interaction between two mesons are repulsive, that's why
it is difficult for these states to form bound states. The contribution of each interaction term to the potential of the system shows that
the kinetic energy term provides attractive interactions, while the Coulomb term and the color-magnetic term provide repulsion,
which decrease the attraction of the kinetic energy term, and make the total weak attraction or repulsion of the system.
(3) By comparing the results of two quark models, the energy of the meson-meson configuration is almost the same in ChQM and QDCSM,
because the $\sigma$ meson exchange of the ChQM is inoperative and the quark delocalization of QDCSM is close to $0$ in the systems
of full-heavy quarks. In the diquark-antidiquark configuration, the energy of QDCSM is smaller than that of ChQM, due to the color
octet symmetry of the diquark and antidiquark, which makes the quark delocalization work in QDCSM, and leads to lower energy in this model.
Nevertheless, the conclusions are consistent in these two quark models.

We study the full-heavy tetraquarks in two structures in this work. To find out more bound states or resonance states, we will explore
the systems with other structures and do the coupling of the different structures. In addition, to confirm the existence of the full-heavy
resonances, the study of the scattering process of the corresponding open channels is needed. In our work, the mass range of these full-charm
resonances is between $6.3$ GeV to $7.4$GeV, and the masses of these full-beauty resonances are around $19.25$ GeV.
The quantum numbers of these possible resonances are $IJ^{P}=00^{+}$, $01^{+}$, and $02^{+}$.
All these full-heavy resonance states are worth searching in the future experiments.

~

\acknowledgments{This work is supported partly by the National Science Foundation
of China under Contract Nos. 11675080, 11775118 and 11535005}.

\end{document}